\documentclass{ifacconf}

\usepackage{graphicx}      
\usepackage{natbib}        
\usepackage{amsmath}
\usepackage{amsfonts}
\usepackage{comment}
\usepackage{float}
\usepackage{todonotes}
\usepackage{subcaption}
\usepackage{caption}
\usepackage{cuted}
\usepackage{siunitx}
\usepackage{placeins}

\newtheorem{theorem}{Theorem}

\newtheorem{lemma}{Lemma}
\newtheorem{assumption}{\textbf{Assumption}}

\begin{document}
\begin{frontmatter}

\title{KalMRACO: Unifying Kalman Filtering and Model Reference Adaptive Control for Robust Control and Estimation} 

\thanks[footnoteinfo]{This work is funded by the European Union through the INESCTEC.OCEAN Center of Excellence in Ocean Research and Engineering (Project number 101136903)}

\author[SINTEF]{Lauritz Rismark Fosso} 
\author[NTNU]{Christian Holden}
\author[SINTEF]{Sveinung Johan Ohrem}

\address[SINTEF]{SINTEF Ocean, Dept. of Energy and Transport, Trondheim, Norway (E-mail: lauritz.fosso@sintef.no; sveinung.ohrem@sintef.no)}
\address[NTNU]{Dept. of Mechanical and Industrial Engineering, Norwegian University of Science and Technology, Trondheim, Norway (E-mail: christian.holden@ntnu.no)}

\begin{abstract}                
A common assumption when applying the Kalman filter is a priori knowledge of the system parameters. These parameters are not necessarily known, and this may limit the real-world applicability of the Kalman filter. The well-established Model Reference Adaptive Controller (MRAC) utilizes a known reference model and ensures that the input-output behavior of a potentially unknown system converges to that of the reference model. We present KalMRACO, a unification of Kalman filtering and MRAC leveraging the reference model of MRAC as the Kalman filter system model, thus eliminating, to a large degree, the need for knowledge of the underlying system parameters in the application of the Kalman filter. We also introduce the concept of blending estimated states and measurements in the feedback law to ensure stability during the initial transient. KalMRACO is validated through simulations and lab trials on an underwater vehicle. Results show superior tracking of the reference model state, observer state convergence, and noise mitigation properties.
\end{abstract}


\end{frontmatter}

\section{Introduction}
The Kalman Filter,~\citep{kalman1960new, kalman_new_1961}, is arguably one of the most important developments in the domain of control theory, and its application has contributed to technological leaps in fields such as robotics~\citep{urrea2021kalman}, dynamic positioning of ships~\citep{balchen1980dynamic} and economics~\citep{athans1974importance}.

The Kalman filter combines direct measurements with a model of the system dynamics to arrive at an estimate of the system states. Under the assumption of linear dynamics and additive zero-mean gaussian noise, the Kalman filter is an optimal solution to the filtering problem, i.e., the problem of determining a signal value from an incomplete and/or noisy observation. For systems with nonlinear dynamics there exist extensions of the Kalman filter framework, at the cost of optimality. The most notable being the extended Kalman filter, where the nonlinear system dynamics are linearized at each iteration, and the unscented Kalman filter, which improves upon the extended Kalman filter by using a sampling approach for estimating of the covariance and mean of the state vector.~\citep{julier_unscented_2004}.

A common assumption when applying the Kalman filter and its derivatives is a priori knowledge of the underlying system model parameters. Particularly, the state transition variable, the input to state variable, and the state to output variables are assumed known. In a real world application it can be impractical or even impossible to obtain these variables, thus limiting the applicability of the Kalman filter~\citep{bulut2012kalman, rocha_robust_2021}.



Uncertain models are a familiar challenge in control theory, and one approach to address the uncertainty is through adaptive control. A central method in adaptive control is model reference adaptive control (MRAC), which has seen steady interest since its inception in 1958~\citep{whitaker_design_1958}. Traditionally, the MRAC scheme uses an open loop reference model containing the desired dynamics of the closed loop system. A controller is designed such that the response of the closed loop system matches that of the open loop reference model. Some modifications to the original MRAC scheme exist, e.g., ~\citet{lee_error_1997} proposed inserting an error feedback term into the reference model leading to improved robustness to nonlinearities. \citet{gibson_adaptive_2013} derived a novel MRAC scheme (which they named Closed-loop Reference Model, CRM) that utilized the states of the reference model in the feedback control loop in an observer-like fashion. \citet{ohrem_controller_2018} had a similar approach, but instead incorporated a Luenberger observer with a copy of the reference model dynamics to provide a state estimate for the controller (the authors called this method Model Reference Adaptive Controller and Observer, MRACO). This approach was extended to MIMO systems in~\citet{ohrem2021adaptive}, and a modified version including an integral term was validated in field trials in~\citet{ohrem_application_2024}. \citet{faramin_modified_2024} proposed a modification to the work of \citet{ohrem_controller_2018}, along with an extension to MIMO systems. They proposed an alternative observer model, and an extra adaptive feedback term. This removed the necessity a parameter upper-bound present in~\citet{ohrem_controller_2018}; however, it was assumed that the system was minimum phase, which is somewhat restrictive.

In~\citet{ohrem_controller_2018} a high observer gain is required to guarantee stability during the initial transient, before any adaptation occurs, when the system model and the reference model exhibit their greatest variation. However, high observer gains limit the practical usability of observers when measurement noise is present, as high gains also amplifies the noise. This contributes negatively to the adaptation, which again leads to unwanted (and potentially unstable) behavior. 

\begin{figure}[htbp]
    \centering
    \includegraphics[width=\linewidth]{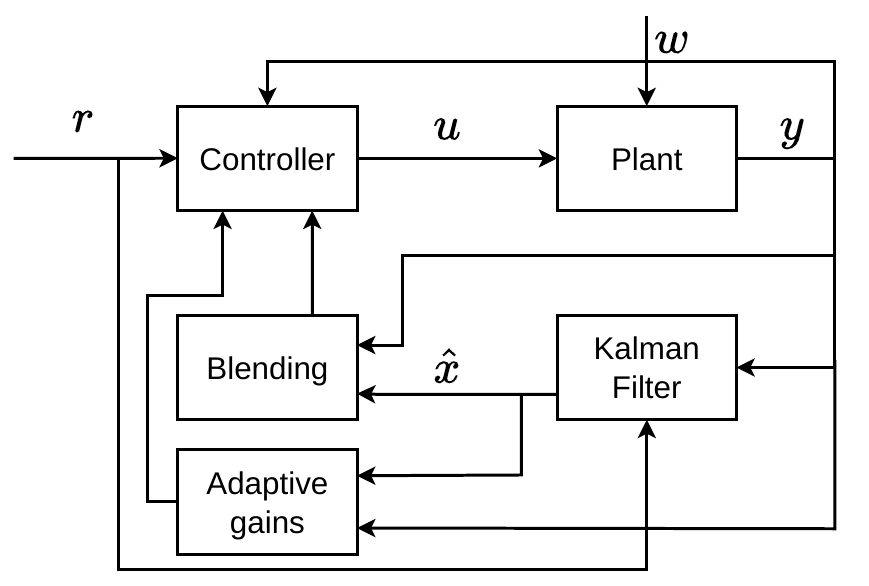}
    \caption{Illustration of KalMRACO}
    \label{fig:MRACO_y}
\end{figure}

In this work, we present, to the best of our knowledge, a new approach combining the optimality of the Kalman filter with the robustness of the MRAC scheme in a new combined method we have named KalMRACO. The main idea is to use the reference model of the MRAC scheme as the system model in a conventional Kalman filter and use the estimated states of the Kalman filter as inputs to the controller. We present an initial, scalar design where the Kalman filter outputs are blended with the system outputs which addresses the stability issue during the initial transient, allowing the time-varying Kalman filter gain to be used instead of the constant Luenberger gain. This removes the high observer gain requirement of~\citet{ohrem_controller_2018}. Furthermore, our proposed scheme removes the need for accurate system models in the Kalman filter algorithm, simplifying conventional Kalman filter implementation. The suggested scheme is depicted in block diagram form in Figure~\ref{fig:MRACO_y}. Simulation results, and results from a real-world implementation on an underwater vehicle support the theoretical results and verifies the proposed scheme.

The key contributions of this paper and the KalMRACO concept are listed below.
\begin{enumerate}
    \item Eliminating the need for precise system models in conventional Kalman filter design by leveraging the MRAC reference model as the Kalman filter system model.
    \item Introducing the concept of blending estimates with measurements in the feedback loop to mitigate initial transient stability issues of existing methods thus allowing the selection of any observer gain $L > 0$, including the time-varying Kalman gain.
    \item Real-world trials on an underwater vehicle demonstrating the applicability and advantages of the proposed scheme.    
\end{enumerate}

\section{Problem Description}
\subsection{System, Reference Model and Observer Description}
Consider the scalar system
\begin{subequations} \label{eq:sys}
\begin{align}
    \dot x &= ax + b(u - w ) \\
    y &= x
\end{align}
\end{subequations}
 where $u$ is the control input, $w$ represents a constant but unknown disturbance, and $a$ and $b$ are unknown.

The following assumptions apply:
\begin{assumption}
    The sign of $b$ is known.
\end{assumption}
\begin{assumption}
    $a$ and $b$ are constant.
\end{assumption}
The reference model of a traditional MRAC scheme can be defined as
\begin{equation}
    \dot x_m = -a_mx_m + b_mr
\end{equation}
where $r$ is a bounded reference signal, and $a_m$ and $b_m$ are design parameters. We define $k^*, l^*$ such that $a_m =bk^* - a$ and $b_m = l^* b$. Since we assume $a$ and $b$ are unknown, we do not know $k^*$ and $l^*$ and hence, it is necessary to replace these with their estimates $\hat{k}$ and $\hat{l}$. The adaptation laws driving the evolution of $\hat{k}$ and $\hat{l}$ will be defined shortly.

Now consider the observer
\begin{equation}
    \dot{\hat{x}} = -a_m \hat x + b_m r + L(t)(y - \hat x)
\end{equation}
 where $L(t)$ is updated according to the Kalman-Bucy equations \citep{kalman_new_1961} 
\begin{subequations} \label{eq:kalman_bucy}
    \begin{align}
        L &= P R^{-1} \\
        \dot{P} &= Q - a_mP - P a_m - L R L^\top \label{eq:ricatti}
    \end{align}
\end{subequations}
where $R$ and $Q$ are the variance of measurement and process noise, respectively. 

\subsection{Control Objective and Limitations of Existing Method}
The control objective is for $x$ to track $x_m$, i.e., $\lim_{t \to \infty} x = x_m$. 
To include the Kalman filter, the control design cannot impose restrictions on the choice of parameter $L(t)$, as this term is decided by the additive process noise $q=  \mathcal{N}(0,Q)$ and measurement noise $v = \mathcal{N}(0,R)$ that is present in the system.

To motivate our main contribution we would like to address a drawback of unfiltered state feedback, i.e., feeding back the measurement directly to the controller without passing it through, for instance, a Kalman filter. The following Lemma addresses this situation.

\begin{lemma}\label{thm:1}
Let the state variables be
\begin{align}
    e_1 &= \hat{x} - x_m \label{eq:e1}\\
    e_2 &= x - \hat{x} \\
    \tilde k &= \hat k - k^* \\
    \tilde l &= \hat l - l^* \\
    \tilde w &= \hat w - w . \label{eq:wtilde}
\end{align}
Then, the input 
\begin{equation}
    u = - \hat k y + \hat l r + \hat w
    \label{eq:u_with_y_input}
\end{equation}
with the adaptive laws satisfying
\begin{align} 
    \dot{\hat k} &=  \mathrm{sgn}(b) y \gamma_1 m_2 e_2 \label{eq:hatk} \\
    \dot{\hat l} &=  -\mathrm{sgn}(b) r \gamma_2 m_2 e_2 \label{eq:hatl}\\
    \dot{\hat w} &=  -\mathrm{sgn}(b) \gamma_3 m_2 e_2 \label{eq:hatw}
\end{align}
renders the origin of the $e_1,e_2, \tilde k, \tilde l, \tilde w$ system stable in the sense of Lyapunov, and additionally $e_1,e_2$ converge to zero.
\end{lemma}

\begin{pf}
The dynamics of the error variable $e_1$ is
\begin{align}
    \dot{e}_1 &= -(a_m + L) \hat{x} + b_m r + Lx + a_m x_m - b_m r \nonumber \\
    &= -a_m e_1 + L e_2. \label{eq:e_1}
\end{align}
The dynamics of $e_2$ is then
\begin{align}
    \dot{e}_2 &= ax + b( \hat{l}r - \hat{k}y) + a_m \hat{x} - b_m r - Lx + L \hat{x} \nonumber \\
     &= - (a_m +L) e_2 -b \tilde{k} y + b\tilde{l}r + b\tilde w. \label{eq:e_2}
\end{align}
Choose the Lyapunov function candidate
\begin{equation}\label{eq:lyapunov}
    V = \frac{1}{2} m_1 e_1^2 + \frac{1}{2} m_2 e_2^2 + \frac{\vert b \vert}{2 \gamma_1} \tilde{k}^2 + \frac{\vert b \vert}{2 \gamma_2} \tilde{l}^2 + \frac{\vert b \vert}{2 \gamma_3} \tilde{w}^2.
\end{equation}
 
The resulting derivative along the trajectories of the system is then
\begin{align}
    \dot V &= e_1 m_1(-a_m e_1 + L e_2) \nonumber \\ 
    &\phantom{=}\; + e_2 m_2\left(b \tilde{l} r - b \tilde{k} y +b\tilde{w} -(a_m + L) e_2\right) \\
    &\phantom{=}\; + \frac{\vert b \vert}{ \gamma_1}  \tilde{k} \dot{\hat{k}} + \frac{\vert b \vert}{ \gamma_2} \tilde{l} \dot{\hat{l}} +  \frac{\vert b \vert}{ \gamma_3} \tilde{w} \dot{\hat{w}} \nonumber
\end{align}
which, after inserting equations \eqref{eq:hatk}--\eqref{eq:hatw}, becomes
\begin{equation}
    \dot V = -\boldsymbol{e}^\top \boldsymbol{U}(L) \boldsymbol{e} 
\end{equation}
 where 
\begin{equation}
    \boldsymbol{U}(L) = \begin{bmatrix}
        a_m m_1 & - \frac{1}{2} L  m_1 \\
        - \frac{1}{2} L  m_1 & (L + a_m) m_2 
    \end{bmatrix} , 
\end{equation} 
and $\boldsymbol{e} = [e_1 \; e_2]^\top$. Note that $L$ and thus $\boldsymbol{U}$ are time-varying. 

Stability in the sense of Lyapunov is satisfied if
\begin{align}
    \dot V \leq - \delta \|\boldsymbol{e}\|_2^2, 
\end{align}
or, equivalently, 
\begin{align}
    \boldsymbol U - \delta \boldsymbol I \geq 0
\end{align}
for some constant $\delta > 0$. 

Positive definiteness of the matrix $\boldsymbol U - \delta \boldsymbol I$ is guaranteed by guaranteeing the positiveness of the principal minors of the matrix:
\begin{align}
    a_m m_1 - \delta &> 0 \label{eq:minor1} \\
    (a_m m_1 - \delta) ([L + a_m] m_2 - \delta) - \frac{1}{4} L^2 m_1^2 &> 0 . 
\end{align}
We can guarantee \eqref{eq:minor1} by choosing sufficiently large $a_m$ or $m_1$. Let 
\begin{align}
    a_m m_1 - \delta = \hat \delta > 0. \label{eq:minor1b}
\end{align}
Then,
\begin{align}
    &\phantom{\implies} (a_m m_1 - \delta) ([L + a_m] m_2 - \delta) - \frac{1}{4} L^2 m_1^2 > 0 \nonumber \\
    &\implies \frac{1}{4} m_1^2 L^2  - \hat \delta m_2 L - \hat \delta (a_m m_2 - \delta)  < 0 \nonumber \\
    &\implies \frac{1}{4} m_1^2 L^2  - \hat \delta m_2 L < 0 \label{eq:minor2}
\end{align}
since $\hat \delta (a_m m_2 - \delta)$ can be chosen to be non-negative. Thus, \eqref{eq:minor2} can be guaranteed if
\begin{align}
    0 < L < \frac{4 \hat \delta m_2}{m_1^2} = L_{\mathrm{max}}.
\end{align}
The limit can be made arbitrarily large by choosing $m_2$ sufficiently large. 

The solution to \eqref{eq:ricatti} evolves monotonically \citep{dieci_preserving_1996}, it therefore suffices that $\{L(0), L({\infty})\} \subseteq (0, L_{\mathrm{max}})$, where the limit is the solution to the Ricatti equation \eqref{eq:ricatti},
\begin{equation}
 L(\infty) = \frac{-a_m R + \sqrt{a_m^2 R^2 + Q R }}{R} \leq L_{\mathrm{max}} . \label{eq:minor2b}
\end{equation}
We can guarantee $L(\infty) \leq L_{max}$ by choosing $m_2$ sufficiently large. Thus, $L(t) \in (0, L_{\mathrm{max}}) \ \forall \ t \geq0 $.

Hence, by choosing $m_1, m_2, \delta$, and $\hat \delta$ (where $\delta$ can be arbitrarily small) such that \eqref{eq:minor1b} and \eqref{eq:minor2b} are satisfied, $\dot V \leq - \delta \|\boldsymbol{e}\|_2^2 \leq 0$. By \citet[Thm. 8.4]{khalil_nonlinear_2002}, $\boldsymbol{e} \to 0$. \qed



\end{pf}

\section{KalMRACO}
In the ideal case where measurements can be made at a high frequency and without noise, this solution yields good results. However, the motivation behind combining the MRACO scheme with the Kalman filter, is the superior noise rejection properties of the Kalman filter. Feeding back noisy measurements has detrimental effects on the performance, particularly in the adaptation laws, which integrate this noise, resulting in drift in the adapted parameters.
This effect can be observed in the the adaptation law \eqref{eq:hatk}, where under measurement noise $y = x + v$, the expectation of the adaptive law satisfies
\begin{equation}\label{eq:Ekhatdot}
    \mathbb{E}[\dot{\hat{k}}] \propto x(x - \hat{x}) + R
\end{equation}
where the measurement noise variance $R$ presents itself as a bias term, causing growth in $\hat{k}$ even when $x = \hat{x}$. This behavior could be remedied with a leakage term, such as a $\sigma$-modification on the adaptation parameters (see, e.g., \citet[Chapter~8]{ioannou_robust_1996}). However, this does not remedy the fundamental problems that come with noisy feedback, such as increased actuator wear and tear. The $\sigma$-modification also comes at the expense of asymptotic stability. 

\subsection{Blending Control Law}
To address the aforementioned issues we present an alternative approach which combines the usage of the measurement $y$ and the state estimate $\hat{x}$.

This approach requires one additional assumption:
\begin{assumption}\label{ass:bmax}
    The sign of $a$ is known, and if $a > 0$, $a \leq a_{\mathrm{max}}$ where $a_{\mathrm{max}}$ is known.
\end{assumption}

\begin{theorem}[Main result]

Let the control law be given by
\begin{equation}
    u = -\hat{k}(\theta(e_2) \hat{x} + [1 - \theta(e_2)] y) + \hat{l}r + \hat w \label{eq:newu}
\end{equation}
where the blending function $\theta(e_2)$ is given by
\begin{align}
    \theta(e_2) &= \frac{2 \alpha}{1 + e^{(\beta \vert e_2 \vert)}} \label{eq:blending_function}\\
    \alpha &= \left\{\begin{array}{cc}
        1 & \forall \;  a \leq 0 \\
        \frac{a_m}{a_m + a_{\mathrm{max}}} & \mathrm{otherwise} 
    \end{array} \right.
\end{align}
where $\beta > 0$ is a design parameter, 
\begin{align}  
    \dot{\hat k} &=  \mathrm{sgn}(b)\left( [1-\theta(e_2)]y + \theta(e_2) \hat x\right) \gamma_1 m_2 e_2 \label{eq:adaption_laws_2} 
\end{align}
and $\hat l, \hat w$ satisfy \eqref{eq:hatl}, \eqref{eq:hatw}.

Then the origin of the system with states given by \eqref{eq:e1}--\eqref{eq:wtilde} is stable in the sense of Lyapunov. Furthermore, $e_1, e_2$ converge to zero. 
\end{theorem}

\begin{pf}
We start by noting that
\begin{align}
    (1-\theta)y + \theta \hat x = y - \theta e_2
\end{align}
as $y = x$ and $e_2 = x - \hat x$. With $u$ as in \eqref{eq:newu}, the observer error dynamics $\dot{e}_2$ are now given by
\begin{equation}
    \dot{e}_2 = - (a_m +L) e_2 + b\tilde{l}r -b \tilde{k}x  + b \hat{k} \theta e_2 + b \tilde w.
\end{equation}

Let the Lyapunov function candidate be given by \eqref{eq:lyapunov}. The time derivative of $V$ along the trajectories of the system is then given by
\begin{align}
    \dot V &= e_1 m_1(-a_m e_1 + L e_2)  \nonumber \\
    &\phantom{=}\; + e_2 m_2(-(a_m + L) e_2 + b \tilde{l} r -b \tilde{k}x + b \tilde w) \\
    &\phantom{=}\; +  m_2 b \hat{k} \theta e_2^2  + \frac{\vert b \vert}{ \gamma_1} \tilde{k} \dot{\hat{k}} + \frac{\vert b \vert}{ \gamma_2} \tilde{l} \dot{\hat{l}} + \frac{\vert b \vert}{ \gamma_3} \tilde{w} \dot{\hat{w}} . \nonumber
\end{align}
Inserting \eqref{eq:hatl}, \eqref{eq:hatw} gives
\begin{align}
    \dot V &= - m_1 a_m e_1^2 + L e_1 e_2 - m_2 (a_m + L) e_2^2 + m_2 b \hat k \theta e_2^2 \nonumber \\
    &\phantom{=}\; - \tilde k \left(\frac{|b|}{\gamma_1} \dot{\hat k} - b m_2 y e_2 \right)  .
\end{align}
Inserting \eqref{eq:adaption_laws_2} gives
\begin{align}
    \dot V 
    &= - \boldsymbol{e}^\top \boldsymbol{U}_2 \boldsymbol{e}
\end{align}
where we have used that $b(\hat k - \tilde k) = bk^* = a + a_m$ and 
\begin{equation}
    \boldsymbol{U}_2 = \begin{bmatrix}
        a_m m_1 & - \frac{1}{2} L  m_1 \\
        - \frac{1}{2} L  m_1 & m_2(L + a_m - \theta (a + a_m)) 
    \end{bmatrix} . 
\end{equation}

Stability in the sense of Lyapunov is satisfied if
\begin{align}
    \dot V \leq - \delta_2 \|\boldsymbol{e}\|_2^2, 
\end{align}
or, equivalently, 
\begin{align}
    \boldsymbol U_2 - \delta_2 \boldsymbol I \geq 0
\end{align}
for some constant $\delta_2 > 0$. 

Positive definiteness of the matrix $\boldsymbol {U}_2 - \delta_2 \boldsymbol I$ is guaranteed by assuring positiveness of the principal minors of $\boldsymbol U_2 - \delta_2 \boldsymbol I$:
\begin{align}
    a_m m_1 - \delta_2 = \hat \delta_2 > 0 \label{eq:2minor1}
\end{align}
\begin{align}
    \hat \delta_2 ([L + a_m - \theta (a + a_m)] m_2 - \delta_2) - \frac{L^2 m_1^2}{4} > 0 \label{eq:2minor2} . 
\end{align}

We can guarantee \eqref{eq:2minor1} by choosing sufficiently large $a_m$ or $m_1$. 
To guarantee \eqref{eq:2minor2}, we necessitate that 
\begin{align}
    a_m - \theta(a + a_m) &\geq 0 \nonumber 
\end{align}
If $a \leq 0$ (the target system is not unstable), this is satisfied as $\theta \leq 1$. Otherwise, we require
\begin{align}
    \theta &\leq \frac{a_m}{a + a_m} . \label{eq:theta_ineq} 
\end{align}
Since $\theta(e_2) \leq \alpha$, we find that
\begin{equation}
    \alpha = \frac{a_m}{a_{\mathrm{max}} + a_m} \leq \frac{a_m}{a + a_m}
\end{equation}
guarantees \eqref{eq:theta_ineq} when $a > 0$.

Furthermore, there exists an $L_{\mathrm{max}}$ such that \eqref{eq:2minor2} is satisfied if 
\begin{align}
    0 < L < L_{\mathrm{max}}
\end{align}
where $L_{\mathrm{max}}$ can be made arbitrarily large by choosing $m_2$ sufficiently large and $ \alpha = a_m/(a_{\mathrm{max}} + a_m)$. The rest of the proof is identical to that of Lemma \ref{thm:1}. \qed
\end{pf}

Introducing measurement noise $y = x + v$, the expectation of the adaptive law \eqref{eq:adaption_laws_2} satisfies
\begin{equation} \label{eq:Ekhatdot_blended}
    \mathbb{E}[\dot{\hat{k}}] \propto x(x - \hat{x}) - \theta(x - \hat{x})^2+ R(1 - \theta)
\end{equation}
from which it is clear that the bias $R$ that is introduced from the measurement noise is now mitigated by $\theta$.

\section{Results}
Experimental validation of the KalMRACO scheme with input given by~\eqref{eq:newu}, blending function given by~\eqref{eq:blending_function} and update laws given by~\eqref{eq:hatl},~\eqref{eq:hatw} and~\eqref{eq:adaption_laws_2} is conducted on the BlueROV2 Heavy, in both simulation and real-world trials.
\subsection{Simulations}
To gauge the real world performance we simulate using a non-linear second-order model of an underwater vehicle,~\citep{fossen_handbook_2021} with BlueROV2 Heavy parameters drawn from~\cite{skaldebo2023modeling}. Though the vehicle of choice has nonlinear dynamics it has been shown in previous studies that the velocity dynamics can be approximated by linear models on the form given in~\eqref{eq:sys},~\citep{ohrem_application_2024}, thus allowing the application of KalMRACO. We also introduce measurement noise $y = x + v$.
Table \ref{tab:parameters} gives an overview of all parameters used in the simulations. The simulation is carried out in MATLAB R2025a. 
  

\subsection{Simulation 1}
To demonstrate the unwanted noise integration effect previously discussed we first perform a simulation using the controller and adaptation from~\eqref{eq:u_with_y_input} and~\eqref{eq:hatk}--\eqref{eq:hatw}, i.e., we do not use the blending technique introduced in this paper (we set $\alpha=0$, first entry of the $\alpha$ vector in Table~\ref{tab:parameters}). For this scenario we assume $a<0$, i.e., we assume a stable target system. The reference signal in this simulation scenario consists of three alternating steps in the desired surge speed between 0 and 0.2 m/s, ending with 0.2 m/s being sustained until the end of the simulation. The same scenario is also simulated with the KalMRACO scheme utilizing the blending technique ($\alpha =1$, second entry of the $\alpha$ vector in Table~\ref{tab:parameters}).

\begin{table}[htbp]
    \centering
    \caption{Parameters used in simulation and lab trials}
    \begin{tabular}{cccc}
        Parameter & Simulation 1 & Simulation 2 & Lab trial \\ \hline
        $\mathrm{sign}(b)$ & 1 & 1 & 1 \\ 
        $\alpha$ & [0, 1] & 10/11 & 10/11 \\ 
        $\beta$ & 1 & 1 & 1 \\ 
       $Q$ & $5 \times 10^{-7}$ & $5 \times 10^{-7}$ &$5 \times 10^{-7}$ \\
       $R$ & $3 \times 10 ^{-4}$ & $3 \times 10 ^{-4}$ & $3 \times 10 ^{-4}$ \\
       $a_m, b_m$ & [1, 1] & [1, 1] & [1, 1] \\
       $a_{max}$ & N/A & 0.1 & 0.1 \\
       $m_1, m_2$ & [1, 70] & [1, 70] & [1, 15] \\
       $\gamma_1, \gamma_2, \gamma_3$& [50 50 5] & [50 50 5] & [50 50 1]   \end{tabular}
       
    \label{tab:parameters}
\end{table}

\begin{figure}[htbp]
    \centering
    \includegraphics[width=0.8\linewidth]{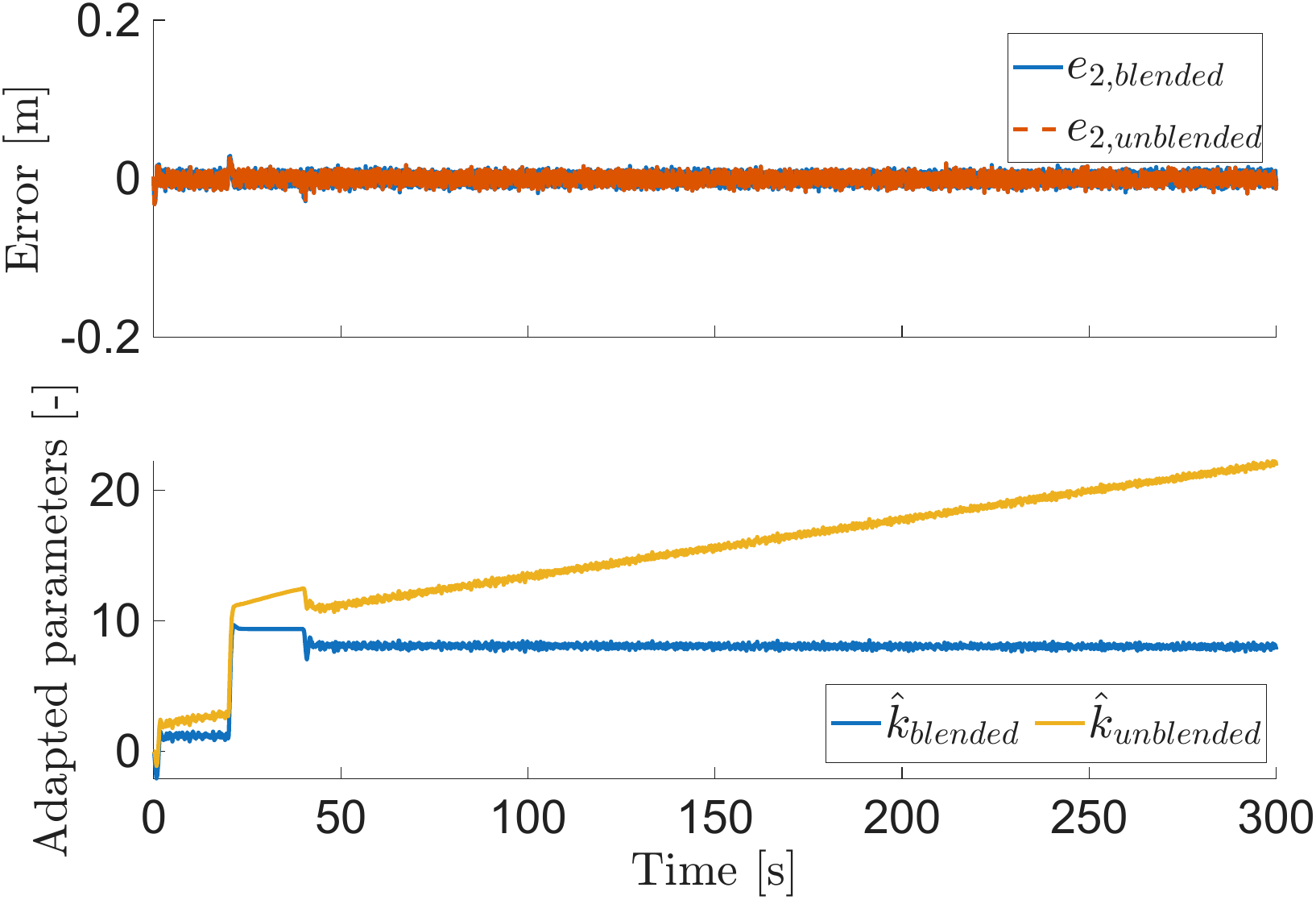}
    \caption{The introduction of blending mitigates the growth in $\hat{k}$ that comes from integration of noise.}
    \label{fig:blending_comparison}
\end{figure}

We see from the bottom plot of Figure~\ref{fig:blending_comparison} that without blending the adapted parameter $\hat{k}$ exhibits undesirable growth only driven by the noise, considering that the underlying error signal is negligible (top plot of Figure~\ref{fig:blending_comparison}). However, $\hat{k}$ does not exhibit this behavior when we apply the KalMRACO scheme and utilize the blending technique, 
as can be seen from the blue line in the bottom plot of Figure~\ref{fig:blending_comparison}. 
This behavior is consistent with the expected dynamics described by \eqref{eq:Ekhatdot} and \eqref{eq:Ekhatdot_blended}.

\subsection{Simulation 2}
With the purpose of demonstrating the robustness capabilities of the proposed scheme we now perform a simulation of KalMRACO on a system assumed to be unstable, i.e., with $a > 0$. We set $a_{max} = 0.1$ which gives $\alpha = 10/11 \approx 0.91$. The objective in the simulation is to track piecewise constant velocity references $x_m$ alternating between 0, 0.2 and 0.3 m/s.

\begin{figure}[htbp]
\centering
\begin{subfigure}[t]{0.48\textwidth}
    \centering
    \includegraphics[width=0.8\linewidth]{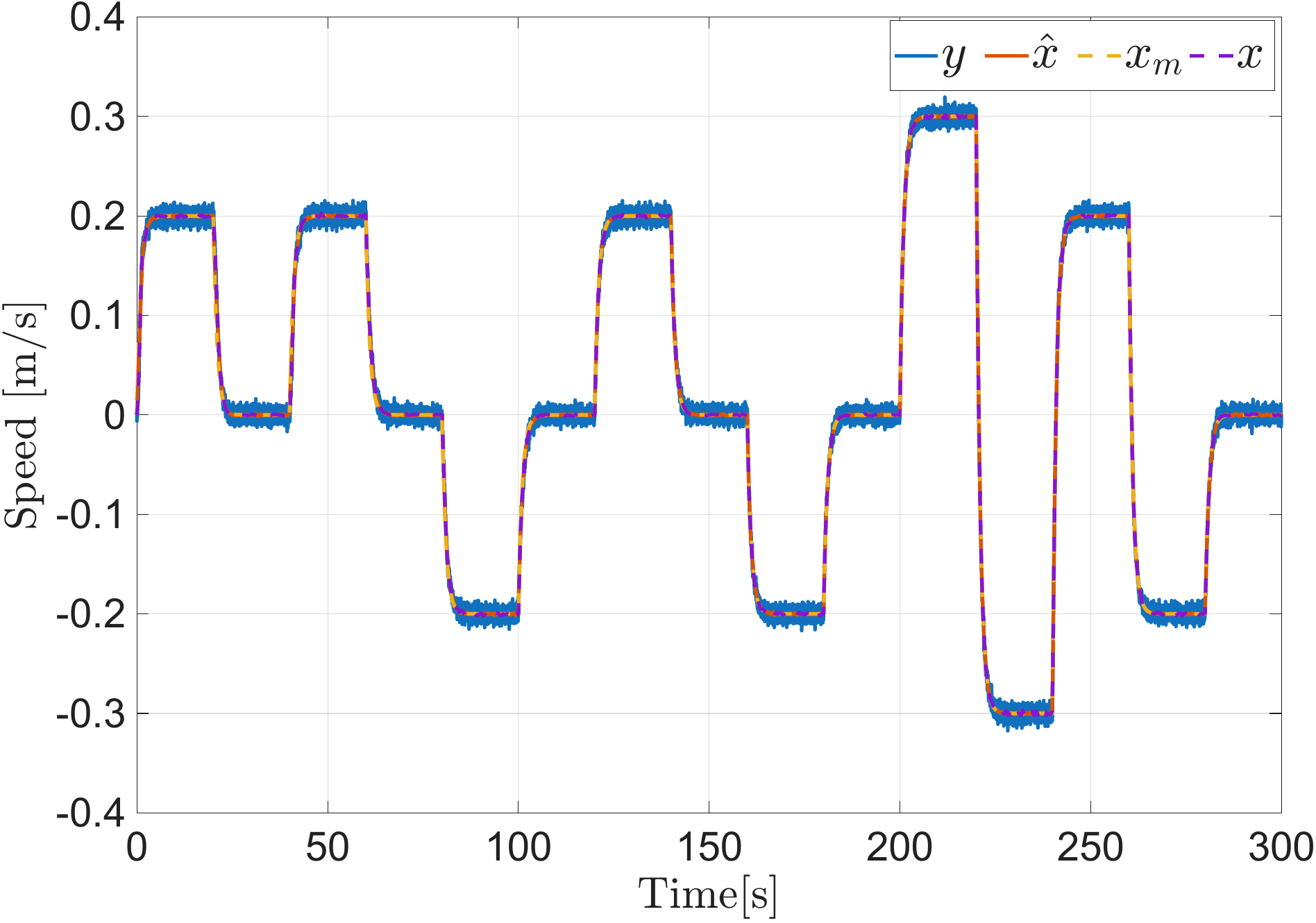}
    \caption{Simulation results of KalMRACO on a system assumed to be unstable. The measured signal $y$ accurately tracks the reference model signal $x_m$ and the Kalman filter state $\hat{x}$ converges to the ground truth state signal $x$, even though this signal is not part of the control or adaptation laws.}
    \label{fig:sim_u}
\end{subfigure}
\begin{subfigure}[t]{0.48\textwidth}
    \centering
    \includegraphics[width=0.8\linewidth]{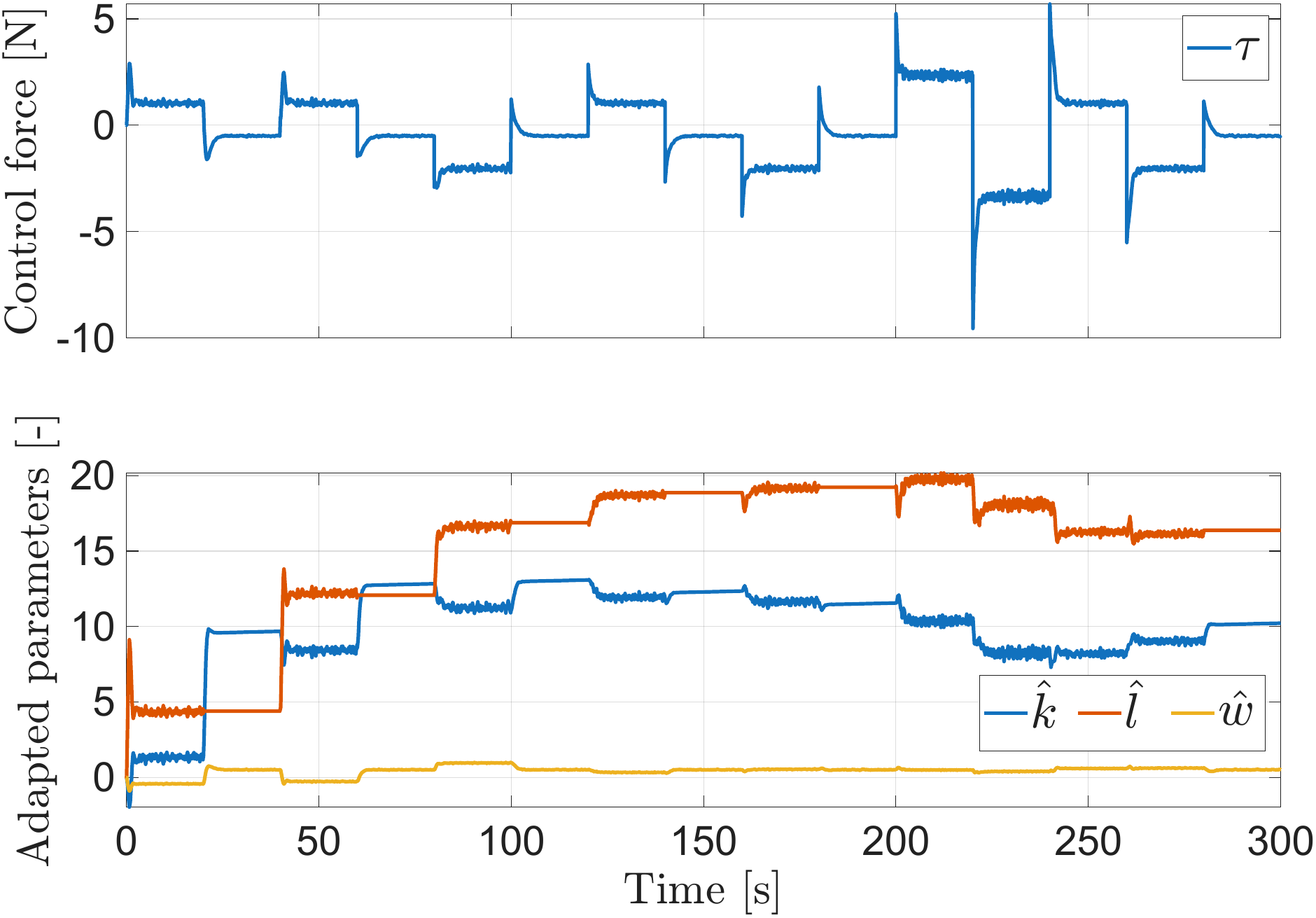}
    \caption{Simulated control force and adapted parameters.}
    \label{fig:sim_tau_param}
\end{subfigure}
\caption{Simulation results of KalMRACO. (a) plots the reference $x_m$, measurement $y$, estimate $\hat{x}$, and ground truth surge speed $x$, while (b) plots the control force $\tau$ and adapted parameters $\hat{k}$, $\hat{l}$ and $\hat{w}$.}
\label{fig:sim_results}
\end{figure}

In Figure~\ref{fig:sim_u} we see that the proposed KalMRACO scheme achieves superior tracking of the references $x_m$ while simultaneously achieving convergence of the Kalman filter estimate $\hat{x}$ to the real value $x$. The adapted parameters in Figure~\ref{fig:sim_tau_param} are well-behaved with no unwanted growth. Note that when the reference $r=0$, there is no adaptation of $\hat{l}$ . Hence $\hat{l}$ remains constant whenever $r\equiv0$; see~\eqref{eq:hatl}. Similar observation can be made with $\hat{k}$; this is because $y$ is small in these instances, which slows adaption (see~\eqref{eq:adaption_laws_2}). 
This can be a challenge when operating in environments with disturbances and is what motivated the introduction of $\hat{w}$ in~\citep{ohrem_application_2024}. Lastly, the calculated control force (Figure \ref{fig:sim_tau_param}, top plot) are feasible and within expected ranges for the BlueROV2 Heavy.

\begin{figure}[htbp]
\centering

\begin{subfigure}[t]{0.48\textwidth}
    \centering
    \includegraphics[width=0.8\linewidth]{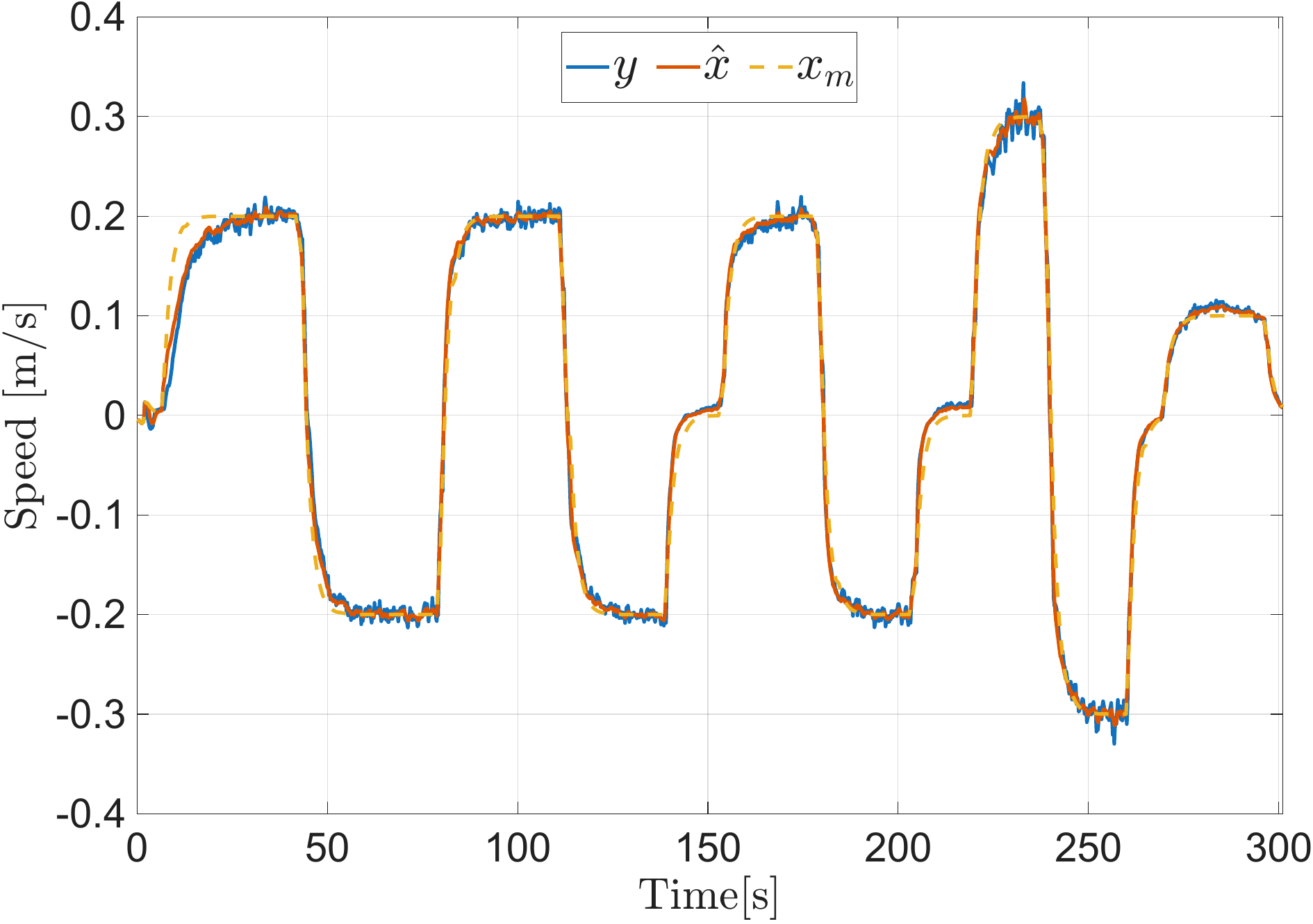}
    \caption{Lab trial results of the KalMRACO scheme.  The measured signal $y$ accurately tracks the reference model signal $x_m$ and the Kalman filter state $\hat{x}$ converges to the measured state.}
    \label{fig:exp_u}
\end{subfigure}
\begin{subfigure}[t]{0.48\textwidth}
    \centering
    \includegraphics[width=0.8\linewidth]{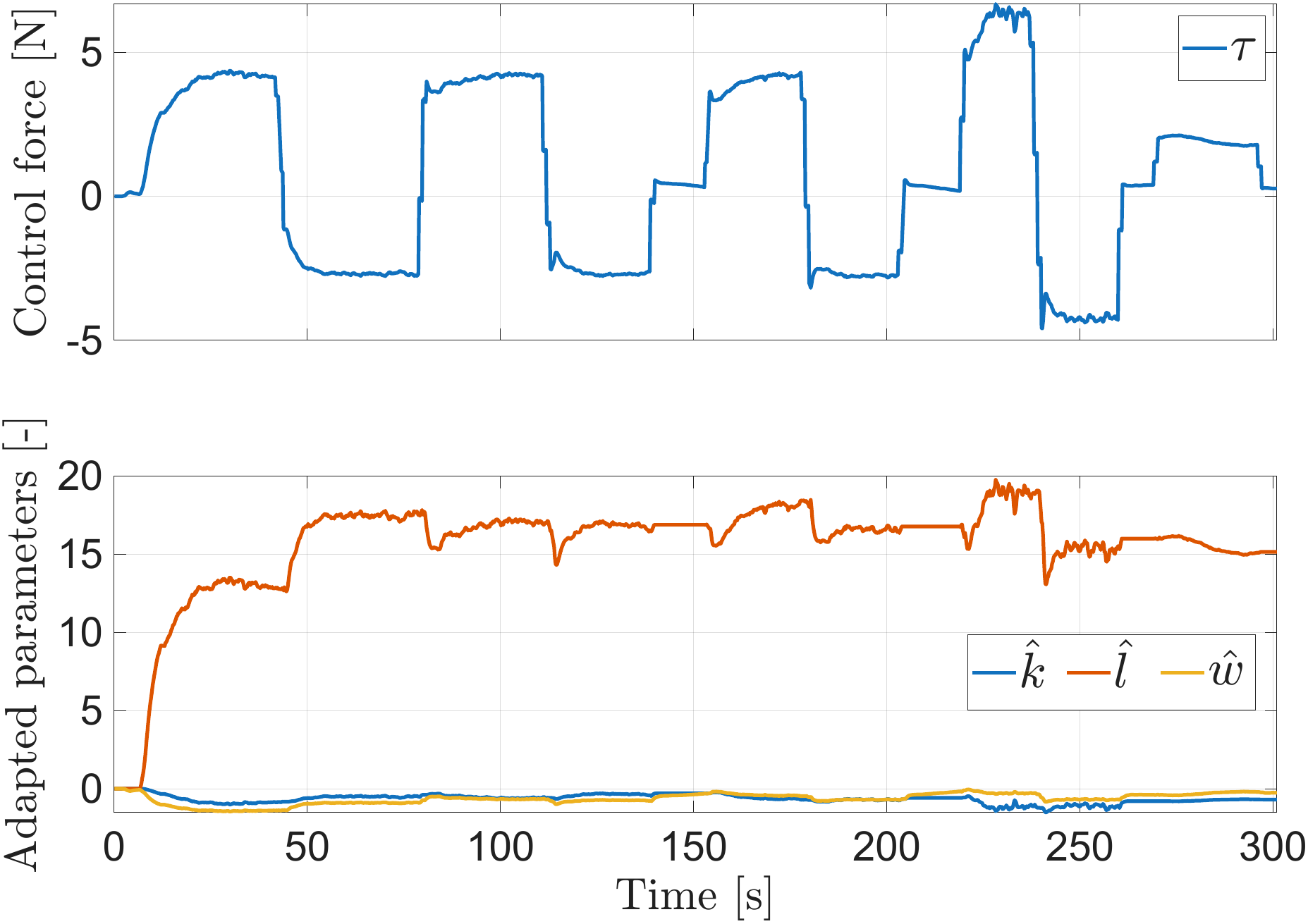}
    \caption{Control force and adapted parameters from lab trials.}
    \label{fig:exp_tau_param}
\end{subfigure}

\caption{Lab trial results of the KalMRACO scheme. (a) plots the model reference $x_m$, measurement $y$, and the Kalman filter estimate $\hat{x}$, while (b) plots the control force $\tau$ and the adapted parameters $\hat{k}$, $\hat{l}$ and $\hat{w}$.}
\label{fig:exp_results}

\end{figure}

\subsection{Lab Trials}
We conduct lab trials with a BlueROV2 Heavy in a pool at SINTEF Ocean in Trondheim, Norway, to further validate the KalMRACO scheme. The parameters used in the lab trials are specified in Table \ref{tab:parameters}. Most parameters are identical to those used in Simulation 2, with the exception of $m_2$, which was reduced to accommodate the increased integration step size used in the real-world trials ($dt = 0.01$ in simulations and $dt =0.1$ in lab trials).

As in Simulation 2 the reference $r$ is piecewise continuous and varies between 0, 0.2 and 0.3 \si{m/s} in the surge direction. Furthermore, we measure the surge speed directly with a Doppler velocity log (DVL) from Water Linked (DVL A50). Figure~\ref{fig:exp_results} reports the trial results. In Figure~\ref{fig:exp_u} we see that the KalMRACO scheme is successful in tracking the reference model state $x_m$ and that the estimated state $\hat{x}$ converges to the measured state $y$ and the reference model state. Further, we note that the estimated state is less contaminated with noise compared to the actual measurement. Also note that the tracking of the reference model state improves for each step in the reference signal. This is as expected, since the adapted parameters are initialized as zero and require some time to adapt. The adapted parameters can be seen in the bottom plot of Figure~\ref{fig:exp_tau_param}. Lastly, the top plot of Figure~\ref{fig:exp_tau_param} shows the applied control force which is smooth due to the blending, mitigating measurement noise.


\section{Conclusions}
This paper presented KalMRACO, a unification of the Kalman filter and MRAC, and introduced blending of estimated and measured states to overcome stability issues during the initial transient. We demonstrated through initial simulation and lab trials that, for scalar systems, the scheme yields excellent tracking of the MRAC reference model state, and convergence of the estimated state to the measured state, while also mitigating noise. Extending the results to higher order systems remains future work


\bibliography{ifacconf}             

\end{document}